\documentclass{moriond}

\def\Journal#1#2#3#4{{#1} {\bf #2}, #3 (#4)}

\def\EPJC{{\em Eur. Phys. J.} C}
\def\JCAP{\em JCAP}
\def\JHEP{\em JHEP}

\def\PLB{{\em Phys.\ Lett.}  B}
\def\PRL{\em Phys. Rev. Lett.}
\def\PR{{\em Phys. Rep.}}
\def\PRD{{\em Phys. Rev.} D}

\def\be{\begin{equation}}
\def\ee{\end{equation}}
\def\bea{\begin{eqnarray}}
\def\eea{\end{eqnarray}}

\def\gsim{\raise0.3ex\hbox{$\;>$\kern-0.75em\raise-1.1ex\hbox{$\sim\;$}}}

\newcommand{\Photo}{}

\begin{document}
\vspace*{4cm}
\title{Supersymmetric models in view of recent LHC data}

\author{W.~Porod}

\address{Institut f\"ur Theoretische Physik und Astrophysik, Uni. W\"urzburg,\\
Campus Hubland Nord, Emil-Hilb-Weg 22, D-97074  W\"urzburg, Germany}

\maketitle\abstracts{We summarize the status of various supersymmetric models in view of
the existing LHC data. A particular focus is on the implications of the measured Higgs mass
on these models which gives important constraints. We consider here minimal and non-mininal 
supersymmetric extension of the Standard Model.
}

\section{Introduction}

The discovery of the Higgs boson at the LHC
\cite{Aad:2012tfa,Chatrchyan:2012xdj} marks one of the most
important milestones in particle physics. Its mass is known
rather precisely: $m_h = 125.09 \pm 0.21$~(stat.)~$\pm
0.11$~(syst.)~GeV~\cite{Aad:2015zhl}. Moreover,
the signal strength of LHC searches in various channels
has been found consistent with  predictions of the Standard Model (SM).
While this completes the SM particle-wise, several
questions still remain open, e.g.\ (i) Is it possible to include the SM
in a grand unified theory where all gauge forces unify? 
(ii) What stabilizes the Higgs mass at the electroweak scale?
(iii) Is there a particle physics explanation of the observed dark matter
relic density? 

Supersymmetry (SUSY) is still one of the best motivated extensions of SM addressing
several of these questions. Consequently, the search for SUSY is among the main priorities 
of the LHC collaborations. Up to now no sign for supersymmetry
or any significant deviation from the Standard Model (SM) prediction has been found, e.g.\
in simplified models bounds on the gluino mass of up to about 2 TeV have been set \cite{Aaboud:2017faq,Sirunyan:2017fsj}
exploiting about 36~fb$^{-1}$ of data in each  experiment. These bounds depend on the spectrum and get
reduced significantly if the spectrum is rather compressed as has been noted early on \cite{LeCompte:2011fh}.

In the minimal supersymmetric standard model (MSSM) the mass of the lighter Higgs boson is bounded to be
below the mass
of the $Z$-boson at tree level implying the need of very large radiative corrections of about 90\%
and larger
as $m_h^2 \simeq m^2_Z + 86^2$~GeV. It has been known for a long time that  large
radiative corrections due to the top-quark and stops, the scalar partners of the top-quark, 
indeed exist as the top-Yukawa coupling is order 1.
This requires that  either the geometric average of the stop masses $M^2_S=m_{\tilde t_1} m_{\tilde t_2}$ is
large and/or the existence of a large trilinear coupling $A_t$ \cite{Djouadi:2005gj} as can be seen
by inspecting  the most
dominant contributions which are given by
\begin{eqnarray}
\Delta m^2_h = 
+ \frac{3 m^4_t}{4\pi^2 v^2} \left[
\ln\left(\frac{M^2_S}{m^2_t}\right) + \frac{X^2_t}{M^2_S}
 \left(1-\frac{X^2_t}{12M^2_S} \right) \right] \,.
\end{eqnarray}
 $X_t=A_t-\mu\cot\beta$ is a measure of the left-right mixing with $\mu$ being the
higgsino mass parameter, $\tan\beta=v_u/v_d$ the ratio of the two vacuum expectation values
and $v^2 = v^2_u+v^2_d= 4 m^2_W/g^2$.

\section{Implication for models with MSSM particle content}

The question, to which extent the observed Higgs mass can be explained within a given
supersymmetric high-scale model and what are the resulting implications
on the spectrum has been investigated by several authors. The main results can be summarized as follows:
in minimal gauge mediated SUSY breaking (GMSB) models one finds $m_{\tilde t_1}\gsim 6$~TeV with
$\tilde t_1$   being the lightest
among the coloured SUSY particles \cite{Ajaib:2012vc}. The main reason is that at the so-called
messenger on finds $A_t=0$ requiring the stops to be heavy.   If this were realized in nature, the LHC at 14 TeV
would not be able to discover SUSY. However,
in case of extended GMSB models one finds
corners in parameter space\cite{Knapen:2016exe} with $m_{\tilde t_1} \simeq m_{\tilde b_1} \gsim 1$~TeV 
which is the mass range explored by the current LHC run \cite{Sirunyan:2017fsj}.
In the constrained MSSM (CMSSM) or slightly extended versions with non-universal Higgs mass parameters
(NUHM-models) the explanation of the observed Higgs mass implies \cite{Baer:2011ab,Kadastik:2011aa,Buchmueller:2011ab}
 $|A_0|\simeq 2 m_0$.  Here $A_0$ and $m_0$ are the trilinear coupling and the
common scalar mass parameter, respectively, at the scale $M_{GUT}$ of grand unification.
Fitting the CMSSM to the Higgs mass taking into account low energy and LHC constraints one
finds that the best fit point \cite{Bechtle:2015nua} has the typical mass scales $m_{\tilde g}, m_{\tilde q} \gsim 2$~TeV, 
$m_{\tilde l_R} \simeq 600$~GeV and $m_{\tilde \chi^0_1} \simeq 450$~GeV. Thus, the up-to now negative search results
is consistent with this part of the parameter which, however, will be  probed by the current and next
LHC runs \cite{Aaboud:2017faq,Sirunyan:2017fsj}. 
Even in more general high scale models with non-universal parameter at $M_{GUT}$ one
typically finds large trilinear couplings\cite{Brummer:2012ns}, e.g.\ 
$|A_0| \simeq (1-3) \max(M_{1/2},  m_{Q_3},  m_{U_3})$.
There is however a problem with large trilinear couplings such as $A_t$ or $A_0$ as they potentially imply 
the existence of a global minimum of the scalar potential which is colour and/or charge breaking. It
 has been shown that large regions of the CMSSM parameter space with $m_h\simeq 125$~GeV 
 are indeed ruled out by color/charge breaking minima  \cite{Camargo-Molina:2013sta}.

High scale models like GMSB or CMSSM imply a rather hierarchical mass spectrum of the supersymmetric
particles giving rise to hard jets and leptons at the LHC in combination with large missing transverse momentum with
only small/tiny SM background. However, in the general MSSM where one can take some parts of the spectrum relatively
compressed leading to substantial reduction of  mass the bounds of various supersymmetric particles \cite{LeCompte:2011fh,Sekmen:2011cz,Arbey:2012bp,CahillRowley:2012kx}.
A particular subclass of the general MSSM are so-called `natural  SUSY' scenarios
\cite{Brust:2011tb,Papucci:2011wy,Hall:2011aa}. Here the basic idea is to take only those 
SUSY particles close the electroweak scale which do give a sizeable contribution to $m_h$ 
in order to avoid a too large fine-tuning of parameters of unrelated sectors and 
to assign to all other particles masses at the multi-TeV scale. In particular, the higgsinos, the partners of the
Higgs bosons, and
the light stop should have masses of the order of a
few hundred GeV. In addition the masses of the gluino and the heavier stop should be close to the TeV scale.
This implies a rather compressed spectrum of the lightest neutralinos and the lightest chargino with
mass differences of $O(1$~GeV$)$
which are rather difficult to detect in direct searches \cite{Barducci:2015ffa} at the LHC. While these
models are interesting from the point of view of fine-tuning they cannot explain the observed
relic dark matter density as the annihilation cross sections of higginos are rather large.
 Moreover, also this class of models requires large
$A_t$ and, thus, gets constrained if one wants to avoid charge and colour breaking minima \cite{Camargo-Molina:2014pwa}.
As already mentioned above, data from the current LHC run imply mass bounds of up to $m_{\tilde t_1} \simeq 1$~TeV
assuming a large mass hierarchy between the stop and the higgsinos \cite{ATLAS:2017kyf}. However, we note
for completeness that even in Natural SUSY the higgsinos might have larger masses  due
to possible existence of the soft SUSY breaking term \cite{Ross:2017kjc} $\mu' \tilde H_u \tilde H_d$ resulting
in higgsino mass of order $\mu+\mu'$.

\section{Extended supersymmetric models}

The requirement of having very large radiative corrections to explain $m_h$ is a hint to go beyond the
MSSM. In non-minimal extensions, the tree-level bound can be pushed to larger values due to the extra
$F$-contributions as in the next to minimal MSSM (NMSSM) \cite{Ellwanger:2006rm} or due to extra 
$D$-term contributions in models
with an enlarged gauge group \cite{Hirsch:2011hg} close to the electroweak scale. 
As  examples we consider $SO(10)$ inspired
left-right symmetric models, which have several virtues: (i) They gives an explanation of the
observed neutrino masses and mixing pattern, (ii) They can explain the conservation of R-parity as $U(1)_{B-L}$ is
a subgroup of $SO(10)$, (iii) The R-sneutrino, the partner of the right-handed neutrino, is a potential
dark matter candidate \cite{Gopalakrishna:2006kr,Arina:2007tm}.
In view of the Higgs mass, taking for example the breaking chain 
\begin{eqnarray*}
SO(10) &\to& SU(3)_C \times SU(2)_L \times SU(2)_R \times U(1)_{B-L} \\
&\to& SU(3)_C \times SU(2)_L \times U(1)_R \times U(1)_{B-L}
\cong  SU(3)_C \times SU(2)_L \times U(1)_Y \times U(1)_\chi
\end{eqnarray*}
on obtains larger tree-level bounds such as \cite{Krauss:2013jva} $m_h \le m_Z^2 + \frac14  g_{\chi}^2 v^2$ 
where $g_\chi \simeq g_Y$ reducing the need for radiative corrections to about 50\% which still is large
but reduces the need for rather large $A_t$ and thus the danger of charge/color breaking minima.
The additional particle content has several phenomenological implications: (i) In particular
in scenarios where a R-sneutrino is the lightest supersymmetric particle (LSP)
 one finds an enhanced lepton multiplicity in the cascade
decays of supersymmetric particles \cite{Hirsch:2012kv}. 
(ii) The existence of a light additional, SM gauge singlet Higgs boson \cite{Hirsch:2011hg,Krauss:2013jva,Krauss:2012ku}. 
(iii) Gauge kinetic mixing and additional $Z'$ decay
modes can significantly alter the LHC bounds on the $Z'$ mass \cite{Krauss:2013jva,Krauss:2012ku}.

\begin{figure}[tbp]
\begin{center}
\includegraphics[width=0.38\textwidth]{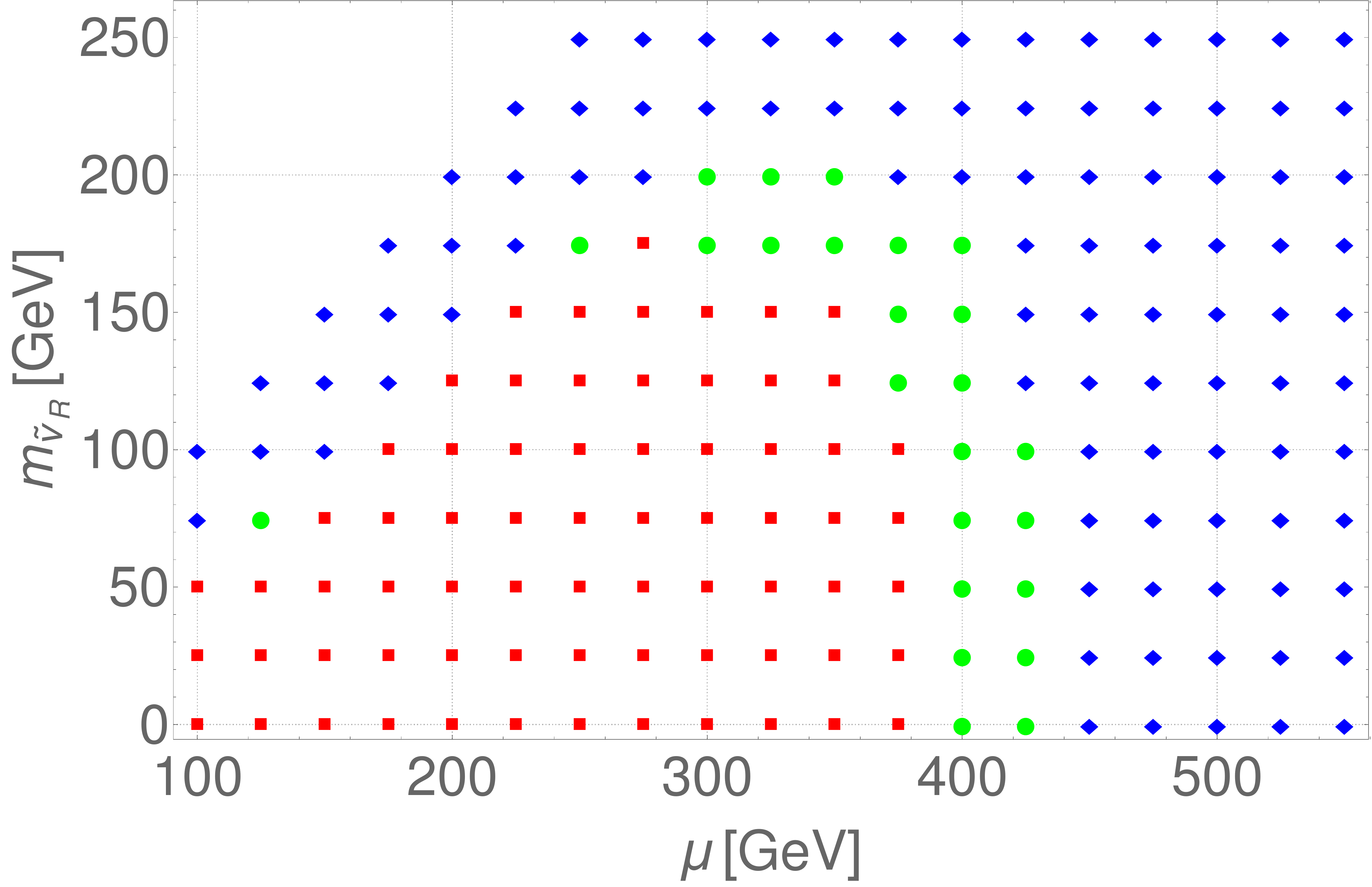}\hspace{14mm}
\includegraphics[width=0.38\textwidth]{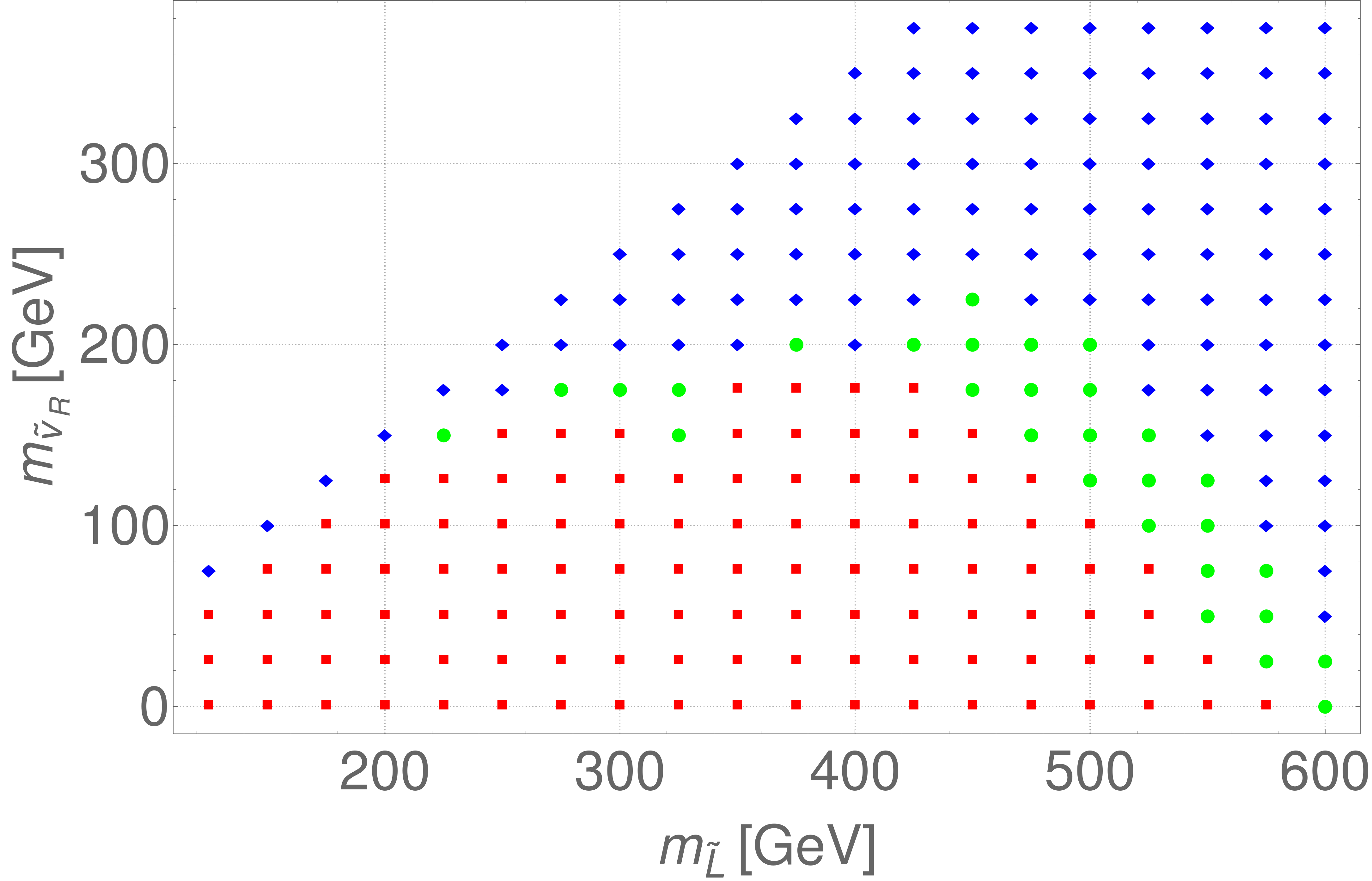} 
\end{center}
\vskip -2mm
\caption{Left: LHC constraints on combinations of $m_{\tilde \nu_R}$ and $\mu$
due to chargino pair production $pp\to \tilde \chi^+_1\tilde \chi^-_1 \to l^+ l^- \tilde\nu_R\tilde\nu_R^*$. 
Right: LHC constraints on combinations of $m_{\tilde \nu_R}$
and $m_{\tilde L}$ due to slepton/sneutrino production in case of an R-sneutrino LSP 
 fixing $\mu=m_{\tilde\nu_R}+25$~GeV, $M_1=M_2=2$~TeV, $m_{\nu_R}=20$~GeV  and $\tan\beta=6$. Red points are excluded, 
blue ones are allowed and
in case of the green ones no conclusive statement can be drawn, within the known
theoretical and experimental uncertainties. See ref.\ $^{34}$ for details.
}
\label{fig:exclusion}
\end{figure}

One might ask if the additional particle content can potentially solve the dark matter problem of Natural SUSY.
In principle, a light right-handed neutrino $\nu_R$ with a mass in the keV range
can do this as in the $\nu$SM \cite{Asaka:2005pn}. We note for completeness
that this is rather difficult to achieve such a scenario in a simple $SO(10)$ scenario. 
If this were the only change, then the LHC phenomenology of Natural SUSY would not change.
 However, it might well be
that the mechanism causing the lightness of the $\nu_R$ implies also that the sleptons and sneutrinos are
rather light. Assuming that a $R$-sneutrino is the LSP  one gets immediately
constraints $\mu$ from higgsino pair production as now the decay 
$\tilde \chi^+_1 \simeq \tilde h^+_1 \to l^+ \tilde \nu_R$ is allowed \cite{Faber2017}. Using
8-TeV and 13 TeV (with an integrated luminosity ${\cal L}=13.9$~fb$^{-1}$) LHC data 
one obtains a bound of about 380 GeV on $\mu$ provided
there is sufficient phase space. In case that also the usual sleptons have masses in the range of a few
hundred GeV, then they are mainly produced via $p p \to \tilde l_L \tilde\nu_L$. In such a scenario one
gets constraints from LHC data on the soft SUSY breaking parameter $M_{\tilde L}$, 
which sets the mass scale of the sleptons,
of up to 580 GeV. We refer to ref.~\cite{Faber2017} for further details. Note, that these bounds apply 
 to any other model containing the corresponding particles.

\section{Conclusions}

Within  the MSSM the explanation of the observed Higgs mass $m_h\simeq 125.1$~GeV  requires large radiative correction. 
This can be either achieved via heavy stops and/or large left-right mixing.
The latter can lead to charge/color breaking minima putting severe constraints on the corresponding parameter
space. Within high scale models such as CMSSM, NUHM or general GMSB, squarks and gluinos have masses in the 1-2 TeV range
in the corresponding parameter regions
which are currently probed at the LHC. If minimal GMSB were realised in nature then the Higgs mass
requires a spectrum of coloured SUSY  particles beyond the reach of LHC at 14 TeV. In generic models with MSSM
particle content the LHC bounds can be substantially reduced if the spectrum is compressed. However, if this realized
in Nature, this will require a quite unusual pattern for supersymmetry breaking as the renormalisation group
evaluation of the underlying parameters yields a quite hierarchical mass spectrum in unified models. 

The relatively large value of $m_h$ might be a hint to go beyond the MSSM as
in non-minimal models additional tree-level contributions to $m_h$ due to F-terms, like in the NMSSM, or due to D-terms,
like in models with extended gauge symmetries, reduce somewhat the need for large radiative corrections.
 We have briefly sketched some important features of $SO(10)$ inspired models. Moreover,
we have shown that in scenarios with an $R$-sneutrino LSP the LHC gives bounds on electroweakly produced
particles of up to 580~GeV.


\section*{References}

\begin{thebibliography}{99}

\bibitem{Aad:2012tfa}
  G.~Aad {\it et al.} [ATLAS Coll.],
  \Journal{\PLB}{716}{1}{2012} 
  [arXiv:1207.7214 [hep-ex]].

\bibitem{Chatrchyan:2012xdj}
  S.~Chatrchyan {\it et al.} [CMS Coll.],
  \Journal{\PLB}{716}{30}{2012}  [arXiv:1207.7235 [hep-ex]].

\bibitem{Aad:2015zhl}
  G.~Aad {\it et al.} [ATLAS and CMS Coll.s],
  \Journal{\PRL}{114}{191803}{2015} 
  [arXiv:1503.07589 [hep-ex]].

\bibitem{Aaboud:2017faq}
  M.~Aaboud {\it et al.} [ATLAS Coll.],
  arXiv:1704.08493 [hep-ex].

\bibitem{Sirunyan:2017fsj}
  A.~M.~Sirunyan {\it et al.} [CMS Coll.],
  arXiv:1705.04673 [hep-ex].

\bibitem{LeCompte:2011fh}
  T.~J.~LeCompte, S.~P.~Martin,
  \Journal{\PRD}{85}{035023}{2012} [arXiv:1111.6897 [hep-ph]].
  
\bibitem{Djouadi:2005gj}
  A.~Djouadi,
  \Journal{\PR}{459}{1}{2008}  [hep-ph/0503173].

\bibitem{Ajaib:2012vc}
  M.~A.~Ajaib {\it et al.}, 
  \Journal{\PLB}{713}{462}{2012} [arXiv:1204.2856 [hep-ph]].
  

\bibitem{Knapen:2016exe}
  S.~Knapen, D.~Redigolo,
  \Journal{\JHEP}{1701}{135}{2017} [arXiv:1606.07501 [hep-ph]].

\bibitem{Baer:2011ab}
  H.~Baer {\it et al.}, 
  \Journal{\PRD}{85}{075010}{2012}  [arXiv:1112.3017 [hep-ph]].

\bibitem{Kadastik:2011aa}
  M.~Kadastik {\it et al.}, 
  \Journal{\JHEP}{1205}{061}{2012} [arXiv:1112.3647 [hep-ph]].

\bibitem{Buchmueller:2011ab}
  O.~Buchmueller {\it et al.},
  \Journal{\EPJC}{72}{2020}{2012}  [arXiv:1112.3564 [hep-ph]].

\bibitem{Bechtle:2015nua}
  P.~Bechtle {\it et al.},
  \Journal{\EPJC}{76}{96}{2016}  [arXiv:1508.05951 [hep-ph]].

\bibitem{Brummer:2012ns}
  F.~Brummer {\it et al.}, 
  \Journal{\JHEP}{1208}{089}{2012}  [arXiv:1204.5977 [hep-ph]].

\bibitem{Camargo-Molina:2013sta}
  J.~E.~Camargo-Molina {\it et al.}, 
  \Journal{\JHEP}{1312}{103}{2013} [arXiv:1309.7212 [hep-ph]].
 
\bibitem{Sekmen:2011cz}
  S.~Sekmen {\it et al.},
  \Journal{\JHEP}{1202}{075}{2012}  [arXiv:1109.5119 [hep-ph]].

\bibitem{Arbey:2012bp}
  A.~Arbey {\it et al.}, 
  \Journal{\PLB}{720}{153}{2013}  [arXiv:1211.4004 [hep-ph]].

\bibitem{CahillRowley:2012kx}
  M.~W.~Cahill-Rowley {\it et al.}, 
   \Journal{\PRD}{88}{035002}{2013}  [arXiv:1211.1981 [hep-ph]].

\bibitem{Brust:2011tb}
  C.~Brust {\it et al.},
  \Journal{\JHEP}{1203}{103}{2012} 
  [arXiv:1110.6670 [hep-ph]].

\bibitem{Papucci:2011wy}
  M.~Papucci {\it et al.}, 
  \Journal{\JHEP}{1209}{035}{2012} [arXiv:1110.6926 [hep-ph]].

\bibitem{Hall:2011aa}
  L.~J.~Hall {\it et al.}, 
  \Journal{\JHEP}{1204}{131}{2012} [arXiv:1112.2703 [hep-ph]].


\bibitem{Barducci:2015ffa}
  D.~Barducci {\it et al.}, 
  \Journal{\JHEP}{1507}{066}{2015}  [arXiv:1504.02472 [hep-ph]].

\bibitem{Camargo-Molina:2014pwa}
  J.~E.~Camargo-Molina {\it et al.}, 
  \Journal{\PLB}{737}{156}{2014} [arXiv:1405.7376 [hep-ph]].

\bibitem{ATLAS:2017kyf}
  The ATLAS Coll.,
  ATLAS-CONF-2017-020.

\bibitem{Ross:2017kjc}
  G.~G.~Ross {\it et al.}, 
  \Journal{\JHEP}{1703}{021}{2017}  [arXiv:1701.03480 [hep-ph]].



\bibitem{Ellwanger:2006rm}
  U.~Ellwanger, C.~Hugonie,
  Mod.\ Phys.\ Lett.\ A {\bf 22} (2007) 1581
  [hep-ph/0612133].
  
\bibitem{Hirsch:2011hg}
  M.~Hirsch {\it et al.}, 
  \Journal{\JHEP}{1202}{084}{2012} 
  [arXiv:1110.3037 [hep-ph]].
  
\bibitem{Gopalakrishna:2006kr}
  S.~Gopalakrishna {\it et al.}, 
  \Journal{\JCAP}{0605}{005}{2006}  [hep-ph/0602027].

\bibitem{Arina:2007tm}
  C.~Arina, N.~Fornengo,
  \Journal{\JHEP}{0711}{029}{2007}  [arXiv:0709.4477 [hep-ph]].
  
\bibitem{Krauss:2013jva}
  M.~E.~Krauss {\it et al.}, 
 \Journal{\PRD}{88}{015014}{2013}  [arXiv:1304.0769 [hep-ph]].
  
\bibitem{Hirsch:2012kv}
  M.~Hirsch {\it et al.}, 
  \Journal{\PRD}{86}{093018}{2012} [arXiv:1206.3516 [hep-ph]].

\bibitem{Krauss:2012ku}
  M.~E.~Krauss {\it et al.}, 
 \Journal{\PRD}{86}{055017}{2012}  [arXiv:1206.3513 [hep-ph]].


\bibitem{Asaka:2005pn}
  T.~Asaka, M.~Shaposhnikov,
  \Journal{\PLB}{620}{17}{2005}  [hep-ph/0505013].

\bibitem{Faber2017} N.~Cerna-Velazco {\it et al.}, 
  arXiv:1705.06583 [hep-ph].  


\end{thebibliography}
\end{document}